\documentclass{optica-article}

\journal{opticajournal} 

\articletype{Research Article}

\usepackage{lineno}

\usepackage{tabularx} 

\usepackage{pifont} 
\usepackage{tikz}
\newcommand{\circled}[1]{\tikz[baseline=(char.base)]{
            \node[shape=circle,draw=gray, fill=gray, text=white, inner sep=0.5pt] (char) {#1};}} 

\begin{document}

\title{Knife-edge removal in schlieren imaging}

\author{Vimod Kumar\authormark{1} and Manish Kumar\authormark{1,*}}

\address{
\authormark{1}Centre for Sensors, Instrumentation and Cyber-Physical System Engineering (SeNSE), Indian Institute of Technology Delhi, Hauz Khas, New Delhi 110016, India
}

\email{\authormark{*}kmanish@iitd.ac.in} 


\begin{abstract*} 
A knife-edge is recognized as a critical component of the schlieren imaging system. This knife-edge serves as a cutoff element, which is required to achieve high sensitivity in the schlieren imaging setup. This article describes a method for totally removing the knife-edge from a normal schlieren imaging system while maintaining the same sensitivity. In this approach, the camera lens's internal aperture acts like the cutoff element making an external knife-edge redundant. This method simplifies the schlieren setup and also reduces the setup cost due to the reduced number of optical or optomechanical elements. We show flow visualization data that clearly demonstrate that schlieren contrast can be obtained by employing a lens's internal aperture as the cutoff. 
\end{abstract*}

\section{INTRODUCTION}
Schlieren is a powerful optical technique for visualizing density gradients in transparent media. This technique closely resembles the Foucault knife-edge test used to test parabolic telescope mirror accuracy \cite{foucault1858description}. However, Foucault paid little attention to the airflow patterns that appeared in the field of view during his knife-edge based optical testing. Around the same time, August Toepler recognized the power of this knife-edge test as a new flow visualization technique and established it as the schlieren technique as we know it today \cite{topler_vibroskopische_1866, topler_optische_1867, settles2001schlieren}. Various configurations for schlieren have been proposed over the years, which include single-lens-based, dual-lens-based, two mirror-based z-type, single mirror-based non-coincident, single mirror-based coincident, and off-axis parabolic mirror-based setups \cite{settles2001schlieren,mercer_optical_2003,schardin_schlierenverfahren_1942, holder1963schlieren,taylor_improvements_1933, barnes_schlieren_1945,zheng_methodology_2022, tp2026study}. 

The knife-edge has been an integral part of all the schlieren configurations mentioned above. A knife-edge is what distinguishes the schlieren from the shadowgraph technique. If we dig further into various ways, we find two techniques that lack the usual knife-edge: focusing schlieren and background-oriented-schlieren (BOS). However, focusing schlieren uses a grid pattern as a cutoff element, which is akin to using multiple knife-edges at the same time. Perhaps BOS is the only truly knife-edge-free schlieren technique where schlieren contrast is obtained by computational image processing \cite{raffel2015background, rabha2025pocket}. If we go back to the seventeenth century, Robert Hooke did perform imaging of hot air rising from a candle where no knife-edge was used. The eye pupil served as the cutoff element to visualize candle plumes diagrammed in \textit{Micrographia} \cite{rienitz1975schlieren, hooke1665micrographia}. Hooke called his technique "the way of concave speculum" \cite{hooke1677lampas, settles2001schlieren}. Perhaps an external knife-edge is fundamental to a schlieren imaging setup because Toepler formally introduced and popularized it in this way. Most contemporary schlieren studies have followed the same procedure and retained the knife-edge cutoff element. Only a few researchers have conducted knife-edge-free schlieren experiments \cite{settles2001schlieren, kannan2020schlieren}. One explanation for this could be a lack of a detailed description of the knife-edge-free schlieren alignment technique. 

In this paper, we present a pure hardware-based method where the knife-edge can be completely replaced by the imaging lens iris, which serves as the essential cutoff element. This approach is a modernized version of Hooke's original technique. The main difference is that Hooke utilized a lens-based, knife-edge-free schlieren system, whereas we demonstrate a mirror-based, knife-edge-free schlieren system. This knife-edge-free schlieren has several advantages, including lower costs and easier alignment while keeping the same level of sensitivity control as knife-edge-assisted schlieren imaging.

\section{METHODS}
\label{sec:methods}
\begin{figure}[htbp]
\centering\includegraphics[width=7cm]{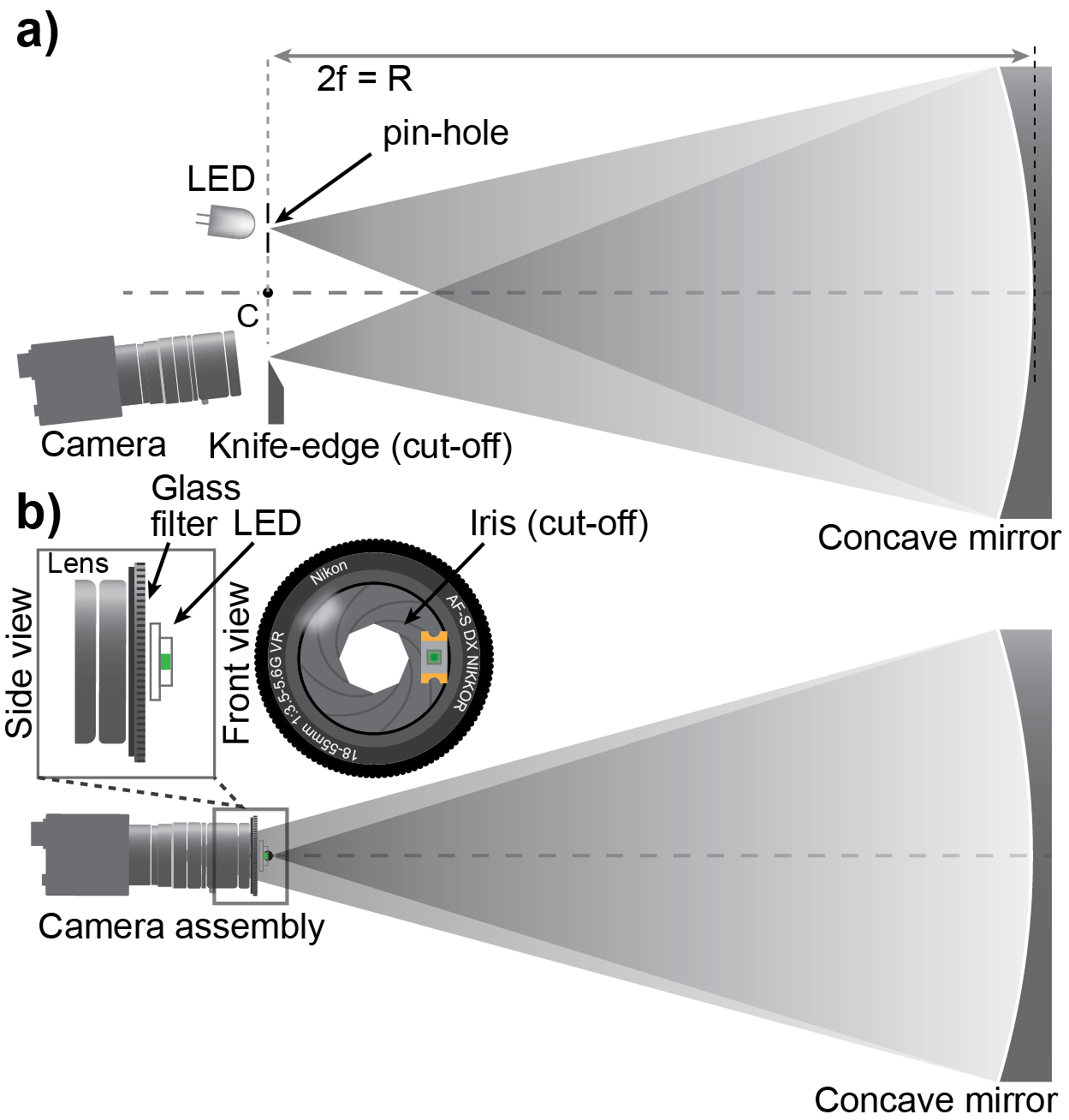}
\caption{Schematics for schlieren imaging systems. a) The conventional knife-edge assisted schlieren imaging system, and b) knife-edge free schlieren imaging system.}
\label{fig:schlieren_schematic}
\end{figure}

We present the alignment method used in a single-mirror schlieren imaging setup. Despite being less common than the two-mirror z-configuration, this setup is very sensitive \cite{gena2020qualitative} and effectively captures the essence of the alignment process, which can also be simply adapted to different schlieren setups. Figure \ref{fig:schlieren_schematic} depicts the schematics for both a typical knife-edge-assisted single-mirror schlieren imaging setup and our knife-edge-free option. The relevant cutoff elements are also highlighted. 

In a typical schlieren configuration, the knife edge must be precisely aligned in both the axial and lateral directions. This alignment is performed concurrently with the light source so that the image of the light source is cut off with the knife-edge. Direct visual inspection is often insufficient to identify the exact cutoff plane, necessitating the positioning of the camera behind the knife-edge. This camera positioning necessitates precise axial and lateral translation. Thus, both the knife-edge and the camera require precise positioning. As a result, removing the knife-edge and accompanying components simplifies the alignment process significantly. We will first look at the component differences between traditional and proposed approaches, followed by their alignment method.

\subsection{List of components} A knife-edge-free schlieren imaging system uses fewer components than a normal knife-edge-aided configuration. The absence of the knife-edge, as well as the associated mounts and stages, accounts for the majority of the difference. Table \ref{tab:list} summarizes the essential components of both configurations, with differences indicated in bold. The knife-edge-free system, with fewer components, costs less. As we'll see in the following section, the knife-edge-free setup alignment is also simpler than its knife-edge assisted counterpart. Supplement 1 contains a list of all of the components used in the demonstrated knife-edge-free schlieren imaging setup, together with their make, model and specifications. 

\begin{table}[htbp]
\caption{Comparison of single-mirror schlieren components with and without knife-edge}
\label{tab:list}
\centering
\begin{tabularx}{\textwidth}{lXX} 
\hline
Component & \textbf{Knife-edge assisted} & \textbf{Knife-edge-free} \\
\hline
Light source & Yes, a small size point-like or thin slit-shaped. & Yes, a small size point-like or thin slit-shaped. \\
Mirror & Yes, a concave mirror. & Yes, a concave mirror. \\
\textbf{Knife-edge} & \textbf{Yes}, a standard knife-edge with its edge oriented parallel to the slit-shaped light source. & \textbf{No}, a knife-edge is not required. \\
Camera & Yes (including lens). Should have tilt and yaw control. & Yes (including lens). Should have tilt and yaw control. \\
\textbf{Height adjustable post} & Yes, \textbf{two} are required. One for the knife-edge and another for the camera. & Yes, \textbf{only one} is required for the camera assembly.\\
Kinematic mount & Yes, required for the mirror to allow for tilt and yaw control.  & Yes, required for the mirror to allow for tilt and yaw control. \\
\textbf{Precision translation stage} & Yes, \textbf{two} or \textbf{three} are required. Knife-edge, camera and potentially light-source need it. & Yes, \textbf{only one} is required for the camera assembly.\\
\textbf{Alignment tools} & Yes, \textbf{three} tools are required. \newline A piece of card or paper (to trace light beam). \newline A viewing screen (to estimate cutoff). \newline A ruler for height measurement & Yes, \textbf{only one} tool is required. \newline  A ruler for height measurement. Other tools may be helpful but are not required (reason described below).\\
\hline
\end{tabularx}
$Note:$ Differences are highlighted in \textbf{bold}.\\
\end{table}

\subsection{Conventional knife-edge assisted schlieren alignment}
For completeness, we begin with a brief description of the alignment technique for traditional knife-edge-assisted single-mirror schlieren imaging. This alignment method is very similar to Foucault's knife-edge test \cite{foucault1858description}. The alignment consists of the following important tasks: i) Setting all optical components to the same centerline height. ii) Locating the plane perpendicular to the optical axis and passing through the mirror's center of curvature, then positioning the light source and knife-edge in this plane so that the image of the light source becomes aligned to the actual knife-edge. iii) Fine-tuning the knife-edge position to ensure a good cutoff. iv) Positioning the camera behind the knife edge and fine-tuning its position to achieve a high schlieren contrast.

The most difficult part of alignment is determining the center of curvature and then placing the image of the light source on the knife-edge cutoff element. This process always requires a visual verification of the light source image position with the help of a small card or paper. These adjustments are often iterative in nature, where adjustment in one optical element (for example, the light source)  necessitates a readjustment in another (for example, the knife-edge). The light source and the knife-edge positions are managed independently. Even if the knife edge and light source are attached to reduce the required number of degrees of freedom during the alignment, finding the center of the curvature plane is based on a manual visual inspection of the light source image falling on the knife edge. This procedure poses practical challenges if the light source is not bright enough or the lab space is brightly lit. Once aligned, the precise positioning of the camera and further fine-tuning of the knife-edge-assisted cutoff percentage add more work. We next describe the alignment method for the knife-edge-free approach, which doesn't need any direct visual inspection for aligning the light source to the cutoff element, thereby greatly simplifying the whole process. 

\subsection{Knife-edge-free schlieren alignment} 
\begin{figure}[htbp]
\centering\includegraphics[width=5cm]{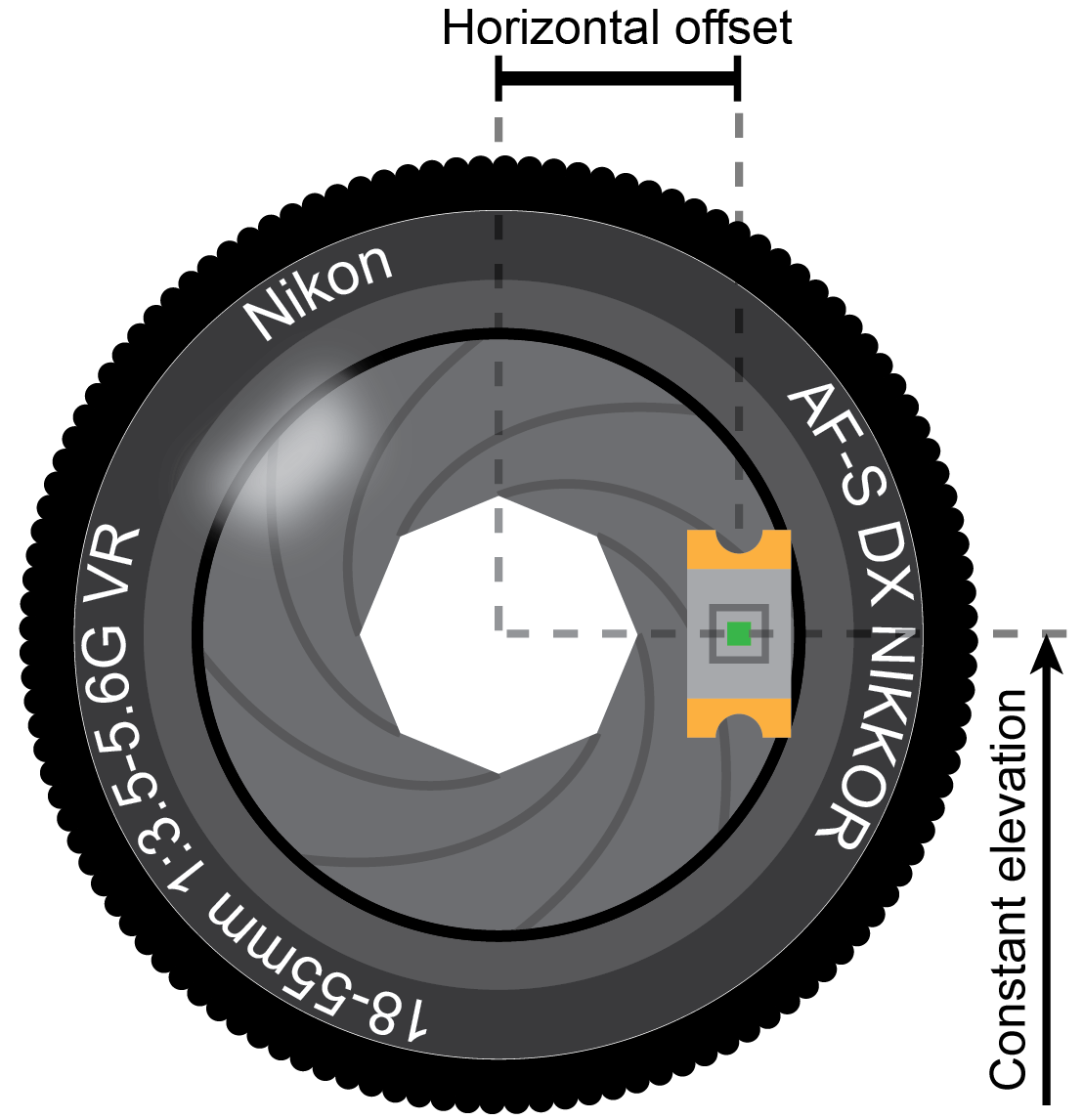}
\caption{A horizontally offset LED placement on camera lens for knife-edge-free schlieren imaging.}
\label{fig:cam-led}
\end{figure}

Our knife-edge-free schlieren imaging system makes use of two major design changes: first, it uses the internal aperture of the camera lens as the cutoff element, and second, it couples the light source and camera lens together. This first change is to remove the knife-edge and utilize a preexisting element as the cutoff. The second change regarding the light source coupling to the camera lens simplifies the alignment. For this, we use wax to attach a small LED light to a blank glass plate filter in front of the camera lens, slightly offset along the horizontal plane, at the same centerline elevation as the iris. Figure \ref{fig:cam-led} shows a schematic for this finished assembly. 

The knife-edge-free schlieren alignment process can be broken into the following broad tasks: i) centerline height alignment, ii) imaging the light source on the iris, and iii) achieving schlieren contrast using the iris edge cutoff. We now describe these tasks in detail. 

\subsubsection {Centerline height alignment}
\begin{figure}[htbp]
\centering\includegraphics[width=12cm]{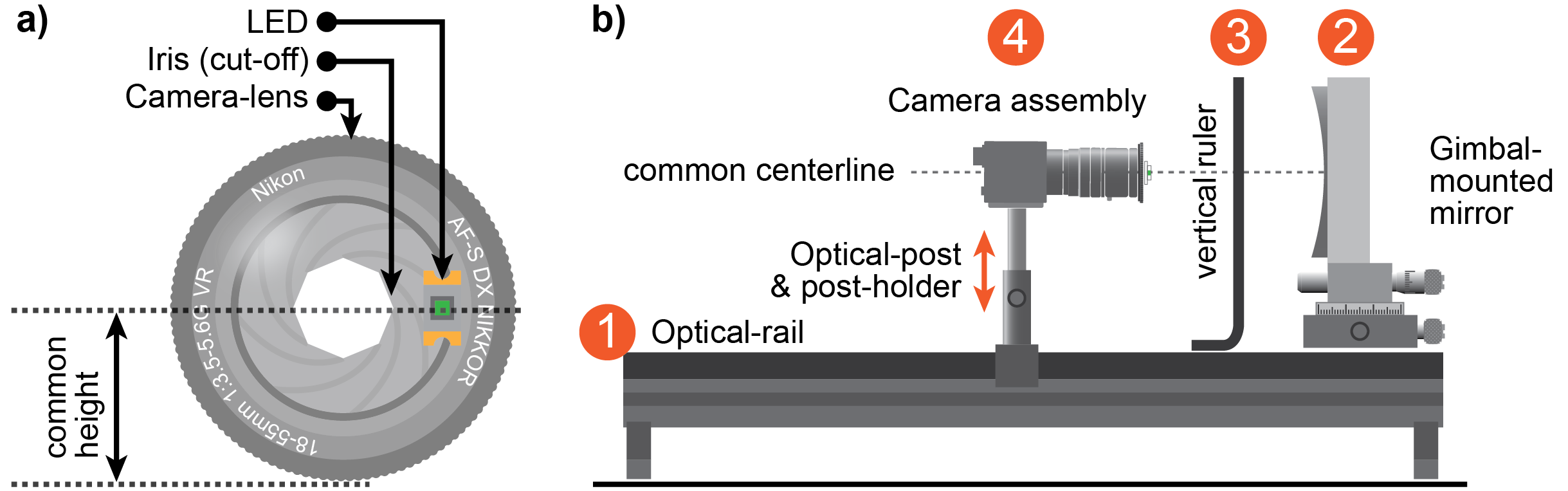}
\caption{Initial assembly of the knife-edge-free schlieren imaging system. All optical elements are aligned at the same centerline height.}
\label{fig:3}
\end{figure}

All optical elements must have the same centerline height with respect to the floor or optical table. The camera and lens iris, which serves as the cutoff element, are already aligned. As previously stated, the light source is fixed at the same centerline height on the lens (see Fig. \ref{fig:cam-led}). As a result, the camera assembly already consists of three key components: the camera, the cutoff element, and the light source, all of which are at a common centerline height. The mirror is vertically mounted on an optical rail (\circled{1} in Fig. \ref{fig:3}) using a gimbal or kinematic mount (\circled{2} in Fig. \ref{fig:3}) with its optical axis along the horizontal plane. We use a vertical ruler (\circled{3} in Fig. \ref{fig:3}) to find mirror's centerline height. Now, the camera assembly is mounted on a height-adjustable optical post and post-holder assembly on top of the optical rail. The camera faces towards the mirror. The camera assembly is brought close to the mirror and using the ruler it is height adjusted to bring it to the same centerline height as the mirror (see \circled{4} in Fig. \ref{fig:3}). The ruler is removed once mirror and camera assembly are at the same centerline height. 

\subsubsection {Imaging the light source in the iris plane} 

The most difficult aspect of system alignment is determining the mirror's center of curvature and aligning the light source and cutoff element with it. However, our knife-edge-free schlieren imaging method eliminates the need for the first half of this step. The alignment task is to directly image the light source on the lens iris. Unlike the knife-edge assisted methodology, our method does not necessitate a direct visual examination of the light source image position. The alignment procedure begins with monitoring the camera's live feed and proceeds through the steps listed below. 

\begin{figure}[htbp]
\centering\includegraphics[width=12cm]{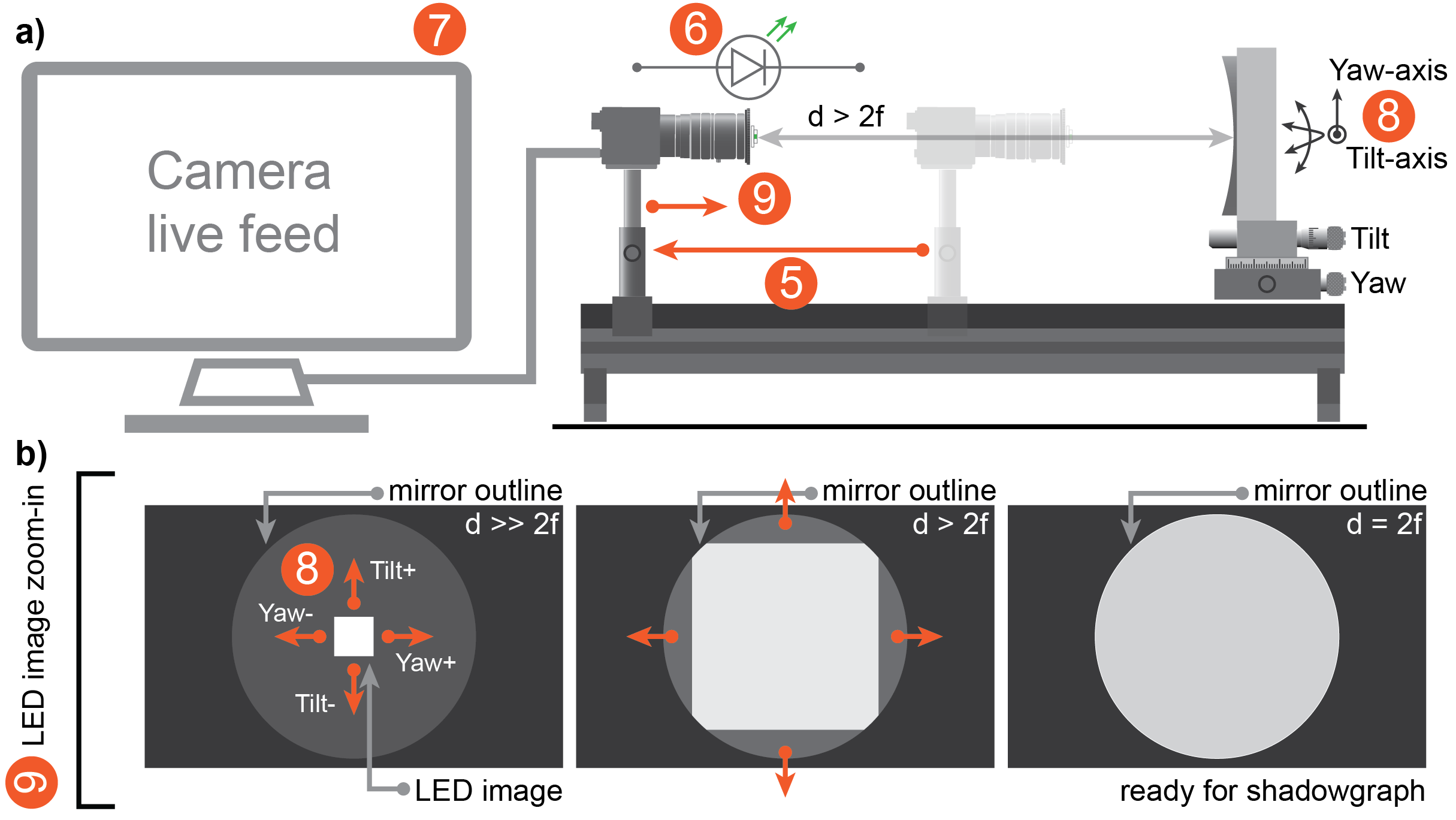}
\caption{Steps for aligning the LED image in the iris plane. a) Schematics representing the sequence of steps and adjustments. b) Illustration of the view from the camera's live feed as the LED image aligns with in the iris plane.}
\label{fig:4}
\end{figure}

\begin{itemize}
    \item The camera assembly is positioned farther than twice the focal length of the mirror, and the LED is powered on (\circled{5} and \circled{6} in Fig. \ref{fig:4}a). The camera live feed shows the mirror (\circled{7} in Fig. \ref{fig:4}a). A small image of the light source should appear within the mirror in the same feed (see Fig. \ref{fig:4}b).

    \item If the image of the light source is absent, the mirror kinematic mount is used to control the tilt and yaw of the mirror (\circled{8} in Fig. \ref{fig:4}a and \ref{fig:4}b) to bring the image of the light source inside the mirror outline. 

    \item The camera assembly is moved closer to the mirror (\circled{9} in Fig. \ref{fig:4}). This action causes the image of the light source to increase in size (Fig. \ref{fig:4}b). If the light source image goes outside the mirror outline, it is brought back in by controlling the yaw (and tilt if required) of the mirror (\circled{8} in Fig. \ref{fig:4}). 

    \item The previous step is repeated until the light source image completely fills the mirror outline. From this position, any further reduction in the distance between the mirror and the camera assembly would decrease the size of the light source image, causing it to underfill the mirror's outline. 

    \item Once the light source image fills the mirror outline, the optimal distance has been realized. At this point, the light source image is physically matched to the lens iris plane.
\end{itemize}

With these steps, the setup is perfectly aligned to perform a shadowgraph. It should be noted that the alignment processes outlined above do not necessitate any direct physical inspection of the light source image. Furthermore, the needed alignment steps were limited to three degrees of freedom: the camera assembly's position on the optical rail , the mirror's tilt, and the mirror's yaw (\circled{8} and \circled{9} in Fig. \ref{fig:4}). This constraint is precisely what makes alignment simple in our method. Up next we describe the steps for performing cutoff and therefore obtain schlieren imaging. 

\subsubsection {Achieving schlieren contrast using the iris edge cutoff} 

The next steps are focused on executing a cutoff to establish the schlieren contrast. Without this cutoff, the resulting images are merely shadowgraphs of any density gradients inserted between the mirror and camera assembly. The following portion describes how to fine-tune the cutoff.

\begin{figure}[htbp]
\centering\includegraphics[width=12cm]{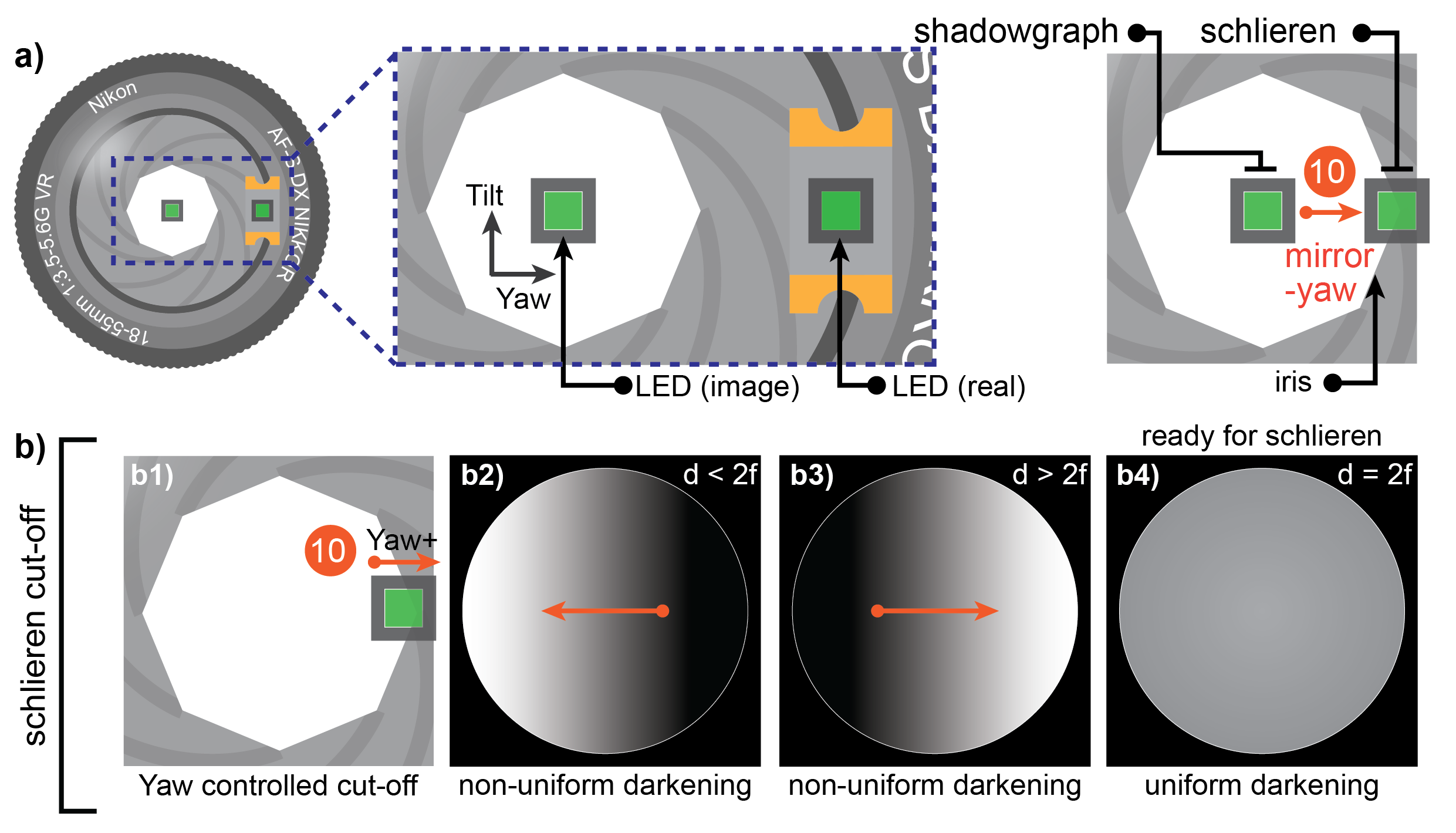}
\caption{Realization of the schlieren contrast using a camera lens iris for cutoff. a) Fine control of the LED image position using mirror's yaw and tilt (on left). Shadowgraph vs schlieren positioning of the LED image (on right). b) The LED image cutoff at iris edge (see b1) causes darkening of the mirror view in camera's live feed. A non-uniform darkening indicates non-ideal distance between camera assembly and mirror (see b2 and b3). A uniform darkening indicates schlieren imaging readiness (see b4).}
\label{fig:5}
\end{figure}

\begin{itemize}     

\item The camera lens iris is set to F8 or a higher aperture value. This corresponds to a narrow iris opening. If closing the iris clips the light source image, the mirror yaw (and tilt, if required) is adjusted to restore light source visibility. This phase may need an iterative approach to vary the iris size and mirror yaw (and tilt, if required) in smaller increments.    

\item The camera lens zoom is controlled by shifting toward longer focal length settings to tightly fill the available field of view with the mirror.    

\item A small solid object, in our case a sharpened pencil, is placed in the area where the flow will be measured, and the lens focus ring is adjusted to sharply image it. The pencil is then removed, and the lens settings for zoom and focus are locked in this position for the remainder of the procedure.    

\item The mirror yaw is gradually adjusted to move the light source image along the horizontal plane towards the edge of the iris opening (\circled{10} in Fig. \ref{fig:5}a). If the camera live feed displays the light-filled mirror darkening unevenly from one side (see Fig. \ref{fig:5}b2 \& b3), it means the iris distance from the mirror is somewhat incorrect in relation to the light source image.

\item If the darkening occurs on the same side as the source image is clipped at the iris edge (Fig. \ref{fig:5}b2), the distance between the camera assembly and the mirror is shorter than necessary. 

\item If the darkening occurs on the other side where the source image is clipped at the iris edge (Fig. \ref{fig:5}b3), the distance between the camera assembly and the mirror is greater than necessary.    

\item The camera assembly distance is accurately adjusted using a micrometer-assisted translation stage to provide a uniform darkening of the light-filled mirror in the camera's live feed.    

\item The cutoff percentage is regulated by fine-tuning the mirror yaw, which grants control over schlieren sensitivity.
\end{itemize}

Throughout the alignment procedure, the camera's exposure time and gain are manually regulated. It may require frequent changes for adequate exposure during zoom and cutoff adjustments. Now the setup is ready for an external knife-edge-free schlieren imaging.

\section{EXPERIMENTAL DEMONSTRATION}
We perform a range of experimental demonstration of the knife-edge-free schlieren imaging. For these experiments we resort to the classical sample of burning candle and thermal plume rising from hands. We also use butane gas flow from a gas lighter as a test sample. The candle plume and butane gas offer strong density gradients covering a large range while hand's plume has a weak density gradient covering small range. Below we present a set of demonstrations adhering to the core alignment strategy described above.

\subsection{Shadowgraph vs schlieren} 
\begin{figure}[htbp]
\centering\includegraphics[width=12cm]{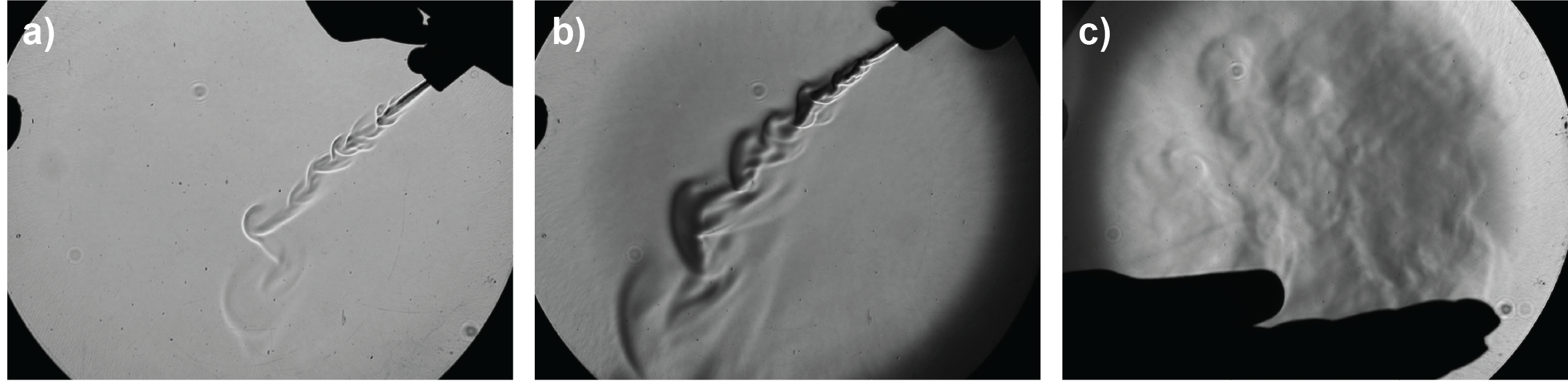}
\caption{Shadowgraph and knife-edge-free high-sensitivity schlieren imaging. Visualization of butane gas flow using a) shadowgraph, and b) knife-edge-free schlieren. c) Knife-edge-free schlieren imaging of a thermal plume from a hand.}
\label{fig:6}
\end{figure}

The biggest difficulty with removing the external knife edge is that one may be performing shadowgraphs rather than schlieren imaging. As a result, our first experimental demonstration will ensure that schlieren can be performed without using a knife edge. We used a 200 mm F/3.3 concave mirror fit inside a kinematic mount. For image capture, we employed a machine vision camera with a normal commercial DSLR lens. We also used a small surface mount LED that was attached directly to the front element of the camera lens. The Appendix contains detailed information about the parts. The assembly was as shown in Fig. \ref{fig:4}a. We utilized a butane gas lighter to discharge butane gas into the detection area near the mirror. The mirror yaw swept the LED image through the lens' internal aperture. This resulted in a cutoff, which marked the change from shadowgraph to significantly improved schlieren contrast for the butane gas flow visualization. Figures \ref{fig:6}a-b depict the results for the shadowgraph and schlieren, respectively. Clearly, the internal lens aperture plays an important role in providing schlieren contrast through LED image cutoff. We also used the Schlieren imaging system to examine the flow of a thermal plume from a hand. This thermal plume offers a small density gradient which is not visible under shadowgraph and necessitates a schlieren imaging based setup \cite{gena2020qualitative}. Figure \ref{fig:6}c shows a camera frame having a clear visualization of this thermal plume. 

\begin{figure}[htbp]
\centering\includegraphics[width=6cm]{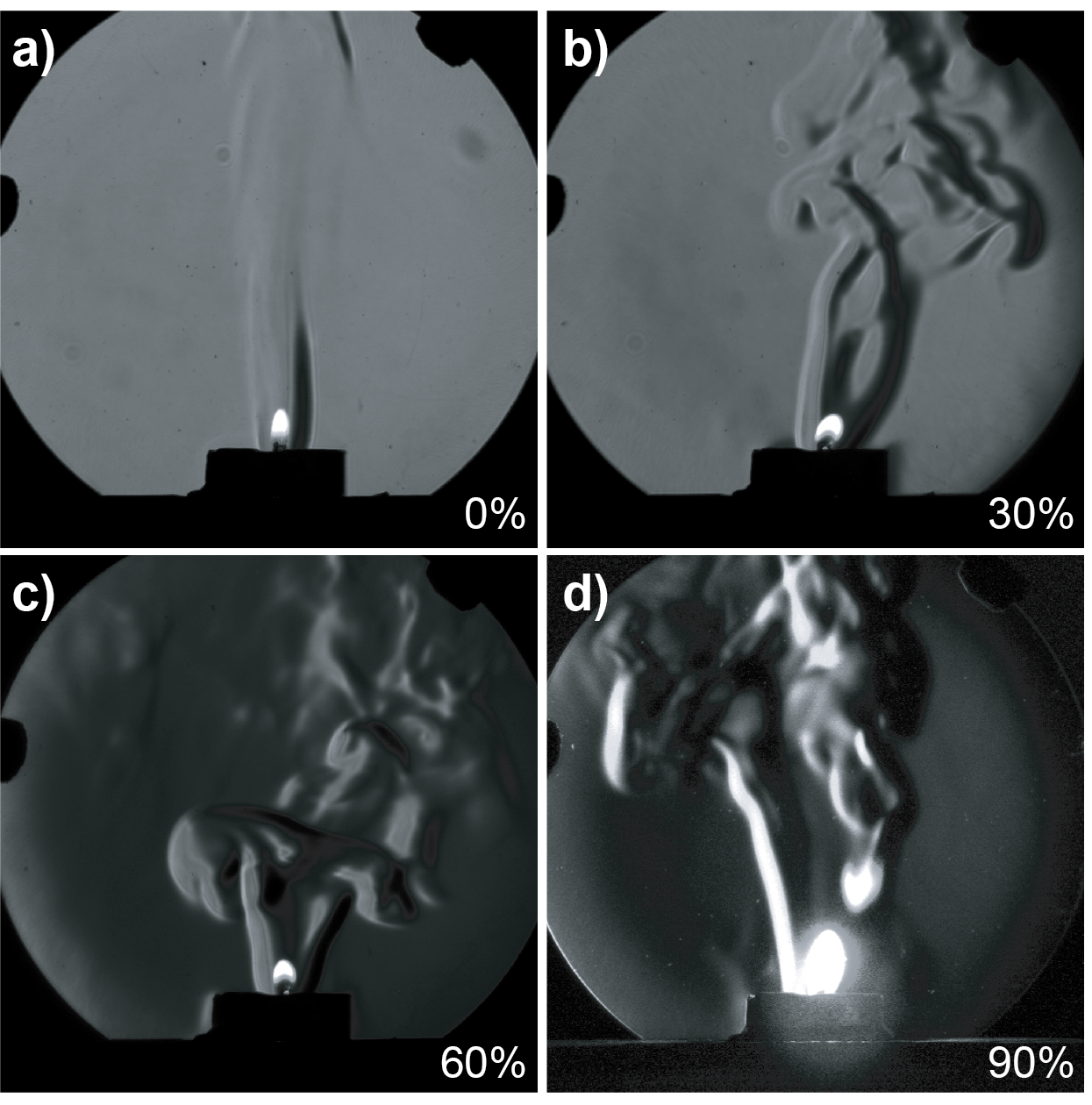}
\caption{Varying degree of the lens-iris assisted cutoff of the LED image demonstrating control on the schlieren contrast and range. The cutoff percentages are approximately a) 0\%, b) 30\%, c) 60\% and d) 90\%. The background intensity levels were matched by exposure control for a-b and c-d pairs.}
\label{fig:7}
\end{figure}
\subsection{Controlling sensitivity and range} A knife-edge is common in schlieren because it allows for different degrees of cutoff of the LED image. The cutoff \% is closely related to the sensitivity and range of schlieren imaging. A higher cutoff increases sensitivity but reduces range, and vice versa. This same level of control is possible with our knife-edge-free schlieren imaging method. We demonstrate this with the classical example of a burning candle. We used the same setup as before to control the mirror yaw to gently run the LED image across the lens iris aperture edge. Figure \ref{fig:7} depicts the flow visualization at various points of this run. As the cutoff \% increases, the background intensity in the camera frame decreases. At the same time, the schlieren contrast increases while the measurable range decreases. At 90\% cutoff, the results show the saturation of the flow zone, indicating a drop in measurable range. As a result, the knife-edge-free schlieren can maintain the same level of control as a normal external knife-edge-assisted schlieren.

\subsection{A minimalist knife-edge-free schlieren setup using a DSLR camera} 
\begin{figure}[htbp]
\centering\includegraphics[width=6cm]{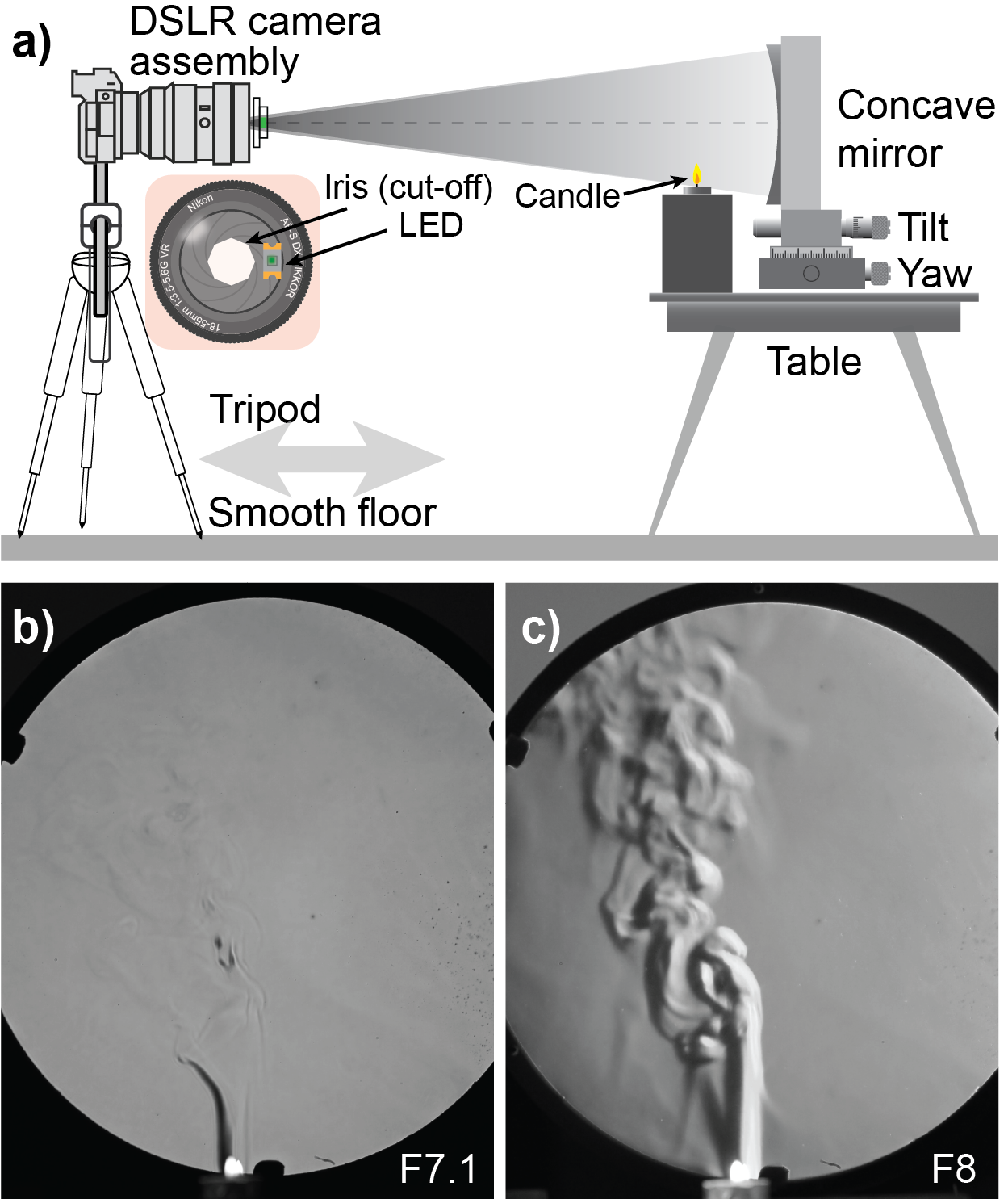}
\caption{A minimal, knife-edge-free schlieren imaging setup. a) The setup uses a tripod-mounted DSLR and a concave mirror. The tripod is slid on the smooth floor to adjust the distance of the DSLR-LED assembly from the mirror. The tilt and yaw control of the mirror enables fine tuning of the schlieren cutoff. b) Shadowgraph of candle plume at F/7.1 aperture value where lens iris does not cutoff LED image. c) Schlieren imaging of a candle plume at an F/8 aperture value where the lens iris acts as a cutoff element for the LED image. Visualization 1 shows this result in a movie format.}
\label{fig:8}
\end{figure}

With experience, it is possible to align a knife-edge-free schlieren setup without a need for an optical rail or precision translation stages. A smooth lab floor can provide a controlled sliding-based fine adjustment to a tripod-mounted DSLR camera to achieve the desirable cutoff. We used this approach in a setup consisting of a 16-inch aperture concave mirror (400 mm focal length, F/4.5). The mirror was mounted in a kinematic mount and placed on top of a table. A tripod helped in centerline height adjustments of a mirror-less camera-LED assembly. Figure \ref{fig:8}a shows the schematics of the complete setup. The large-aperture mirror provided a much better field of view for the candle plume flow visualization. Once aligned, we tried a different approach to switch between shadowgraph and schlieren configurations. We used the dial on the camera to control the lens aperture size. A large iris opening, corresponding to a smaller F-number value (see Fig. \ref{fig:8}b), led to a shadowgraph. On the other hand, a smaller iris size (see Fig. \ref{fig:8}c) led to the schlieren cutoff. This experiment has two effects: first, it makes switching between shadowgraph and schlieren very convenient, and second, it further reinforces that the lens iris acts as the cutoff element to provide a schlieren contrast. 

\bigskip 
\section{DISCUSSION AND CONCLUSION}
The knife-edge is considered deeply tied to schlieren imaging. One cannot envision schlieren imaging without a knife-edge. However, schlieren imaging relies on the knife-edge's role as a cutoff element rather than the knife-edge itself. To achieve schlieren imaging, we confirmed that the role of the cutoff element can be transferred from an exterior knife-edge to an internal lens aperture or lens-iris. To that goal, we developed a comprehensive alignment technique and conducted several experiments to demonstrate that knife-edge-free schlieren imaging is both achievable and convenient. The ability to change between shadowgraph and schlieren by a simple control of the lens aperture size presents an interesting approach. This approach highlights that a lens iris not only acts as a cutoff element but also as a variable cutoff element. This makes the whole process of obtaining a shadowgraph vs. a schlieren heavily reproducible. 

In the section \ref{sec:methods} we have provided details for the knife-edge-free alignment strategy. The proposed approach represented through Fig. \ref{fig:3} to Fig. \ref{fig:5} focuses on having the camera-LED assembly on an optical rail and the concave mirror on a kinematic mount. The core concept in the alignment strategy relies on three degrees of freedom. Tilt and yaw-based two rotational degrees of freedom to the mirror and one translation-based degree of freedom assigned to the camera-LED assembly. However, this is just one of the many possible approaches. The translational degree of freedom could alternatively be assigned to the mirror as well. Similarly, the purpose of the rotational degree of freedom is to enable lateral sweeping of the LED image. This could also be achieved by lateral motion of the camera assembly. Thus, if the camera assembly is provided with a 3-axis precision translation, the alignment task can be completed with a completely static mirror. Our choice of using the mirror tilt and yaw for this fine control was mainly due to the fine adjustment capability of these rotations easily available with a standard mirror mount. 

Although we have described the approach heavily relying on the live camera feed, it is possible to perform the initial alignment manually. If the space is dark and the light source is bright, one may use the closed aperture iris as a screen and form the image of the source directly on the iris. This approach works better for a compact camera lens, as it has a smaller number of glass surfaces before the iris plane is reached. For highly corrected multi-element camera lenses, the amount of back reflection makes it somewhat challenging to see the exact position of the LED image on the iris plane. 

Vibrations play a big detrimental role in schlieren imaging. A vibration in either the mirror or the camera assembly directly changes the cutoff level, leading to a change in the schlieren contrast of the flow. Thus, the vibrations get amplified during a flow visualization experiment. It is crucial to reduce vibrations by ensuring every part is tightly fit and tables or tripods are heavy and stable. For the minimalist setup used in Fig. \ref{fig:8}, we find large tiled floors help with the smooth adjustments of the camera tripod. Similarly, adding more weight to a standard table helps reduce mirror vibrations during the tilt and yaw adjustments. Whenever practically possible, it is suggested to have the mirror and camera assembly mounted on a vibration isolation optical table. 

In conclusion, this paper provides a practical approach to obtain schlieren imaging without a knife-edge. The camera iris-assisted cutoff is both convenient and easily adjustable. We strongly believe that our work will make the knife-edge-free schlieren imaging more accessible to all.

\begin{backmatter}
\bmsection{Funding}
ANRF: SRG/2022/000930

\bmsection{Acknowledgment}
MK acknowledges the Indian Institute of Technology Delhi and Anusandhan National Research Foundation for the startup grants. 

\bmsection{Disclosures}
The authors declare no conflicts of interest.

\medskip

\bmsection{Data availability} Data underlying the results presented in this paper are available at doi: 10.5281/zenodo.18921590 \cite{kumar_2026_18921590}.

\medskip

\bmsection{Supplemental document}
See Supplement 1 for supporting content. 

\end{backmatter}


\bibliography{sample}






\end{document}